\documentclass[useAMS,usenatbib]{mn2e}              
\usepackage{epsfig}
\newcommand{\beq}{\begin{equation}}
\newcommand{\eeq}{\end{equation}}

\newdimen\hssize
\newdimen\hdsize 
\title[Estimating the HI gas fractions of galaxies] 
{Estimating the H  {\sc  i} gas fractions of galaxies in the local Universe}  
\author[Zhang et  al.]{Wei Zhang$^{1}$\thanks{E-mail:   xtwfn@bao.ac.cn},   
  Cheng   Li$^{2,3}$,
  Guinevere   Kauffmann$^{2}$,  
  Hu Zou$^{1}$, 
  Barbara Catinella$^{2}$, 
  \newauthor
  Shiyin Shen$^{3}$, Qi Guo$^{2}$, Ruixiang Chang$^{3}$
  \\  
  ${^1}$  National  Astronomical
  Observatories,  Chinese Academy of  Sciences, Beijing  100012, China
  \\    
  ${^2}$     Max    Planck    Institut     f\"ur    Astrophysik,
  Karl-Schwarzschild-Strasse  1,  85748  Garching, Germany  
  \\  ${^3}$
  Key Laboratory for Research in Galaxies and Cosmology, Shanghai Astronomical Observatory,  Nandan Road 80, Shanghai 200030,
  China 
}

\begin{document}

%\graphicspath{{figs/}}

\date{Accepted ........ Received ........; in original form ........}

\pagerange{\pageref{firstpage}--\pageref{lastpage}} \pubyear{2008}

\maketitle

\label{firstpage}

\begin{abstract}

We use a sample of $800$ galaxies with H {\sc i} mass measurements from the
HyperLeda catalogue and optical photometry from the fourth data release of the
Sloan Digital Sky Survey to calibrate a new photometric estimator of the H {\sc
i}-to-stellar mass ratio for nearby galaxies. Our estimator, which is motivated
by the Kennicutt-Schmidt star formation law, is $\log_{10}(G_{HI}/S) =
-1.73238(g-r) + 0.215182\mu_i - 4.08451$, where $\mu_i$ is the $i$-band surface
brighteness and $g-r$ is the optical colour estimated from the $g$- and $r-$band
Petrosian apparent magnitudes. This estimator has a scatter of $\sigma=0.31$
dex in $\log(G_{HI}/S)$, compared to $\sigma\sim0.4$ dex for previous
estimators that were based on colour alone. We investigate whether the
residuals in our estimate of $\log(G_{HI}/S)$ depend in a systematic way on a
variety of different galaxy properties. We find no effect as a function of
stellar mass or 4000 \AA\ break strength, but there is a systematic effect as a
function of the concentration index of the light. We then apply our estimator
to a sample of $10^5$ emission-line galaxies in the SDSS DR4 and derive an
estimate of the H {\sc i} mass function, which is in excellent agreement with
recent results from H {\sc i} blind surveys. Finally, we re-examine the well-known
relation between gas-phase metallicity and stellar mass and ask whether there
is a dependence on H {\sc i}-to-stellar mass ratio, as predicted by chemical
evolution models. We do find that gas-poor galaxies are more metal rich at
fixed stellar mass. We compare our results with the semi-analytic models of
De Lucia \& Blaizot, which include supernova feedback, as well as the
cosmological infall of gas.

\end{abstract}


\section{Introduction}\label{sec:introduction}

The standard model of galaxy formation posits that galaxies form when gas
cools, condenses and forms stars at the centres of dark matter halos. In
recent years, there has been an explosion of ground- and space-based surveys
that have allowed astronomers to obtain imaging and spectroscopy for samples
of many thousands of galaxies in the local Universe and at high redshifts.
The vast majority of these surveys have probed the rest-frame optical,
ultraviolet or infrared regions of the spectral energy distributions of the
galaxies, and have thus provided important constraints on the properties of
the {\em stars} in these systems. Thanks to these surveys, we have learned
a huge amount about how the stellar masses and star formation rates of
galaxies evolve with time. However, if we are to understand how galaxies form,
we also need to understand how gas is accreted by galaxies and the efficiency
with which that gas is converted into stars.

Our understanding of the cold gas content of galaxies lags considerably behind
our understanding of their stellar populations. In the nearby Universe, new
large surveys such as The Arecibo Legacy Fast ALFA (ALFALFA) Survey of H
{\sc i}, which will detect 25,000 extragalactic H {\sc i} line sources out to
$z \sim 0.06$ using 305m telescope and seven-beam Arecibo L-band Feed Array (ALFA)
\citep{Giovanelli-05}, will do much to redress the balance, but
our poor knowledge of the atomic gas content of galaxies at high redshifts is
likely to persist for many more years. For this reason, there have been several
recent attempts to calibrate colours or emission line equivalent widths as
proxies for the gas-to-stellar mass ratio.

\cite{Tremonti-04} converted star formation surface densities (estimated
from attenuation-corrected H$\alpha$ luminosities) to surface gas mass
densities $\Sigma_{gas}$, by inverting the composite Schmidt law of
\cite{Kennicutt-98}. These indirect gas mass estimates were used (in
conjunction with true gas measurements for a minority of the galaxies) to argue
that the observed correlation between stellar mass and gas-phase metallicity
could not be explained within the context of a closed-box chemical evolution
model. The same technique has been applied to high redshift galaxies by
\cite{Erb-06b} to interpret the redshift evolution in the mass-metallicity
relation, and by \cite{Bouche-07} to argue that high redshift galaxies lie on a
``universal'' star formation relation. Needless to say, the conclusions reached
in these papers are only valid if the same Kennicutt-Schmidt law applies at
high redshifts. 

Because the H$\alpha$ line is not accessible in high redshift galaxies without
near-infrared spectra , there have also been attempts to calibrate
gas-to-stellar mass ratios using optical or optical-infrared colours. The H
{\sc i} gas-to-stellar mass ratio, $G_{HI}/S$, has been found to correlate with
optical (e.g. $u-r$) and optical-NIR (e.g. $u-K$) colours with a typical
scatter of $\sim 0.4$ dex \citep[] [hereafter K04]{Kannappan-04}. Since the
stellar mass of a galaxy can also be estimated from its optical/NIR flux if its
redshift is known \citep[see for example] [hereafter B03]{Bell-03} , these
correlations provide a way of estimating gas masses for large samples of
galaxies where only photometry is available. Although such gas fraction
estimates have large errors for individual galaxies, they may still be useful
for statistical studies. 

In this paper we extend the analysis of K04 by examining the correlation of
$G_{HI}/S$ with additional galaxy properties, in the hope of finding an
estimator of H {\sc i} mass with less scatter. We first demonstrate that we
reproduce the result of K04 if we use the same $G_{HI}/S$ estimator and 
the same sample selection criteria as in K04. We then extend
the analysis by using a larger calibrating sample of galaxies with both
photometry and H {\sc i} masses, and we examine the correlations between
$G_{HI}/S$ and a large variety of parameters. In particular, we show that if we
combine the observed colour with an estimate of the surface brightness of the
galaxy, we can reduce the scatter in our estimates of $G_{HI}/S$ by a
substantial factor. Finally, we apply our best estimator to a large sample of
star-forming galaxies selected from the SDSS DR4. We show that we recover the H
{\sc i} mass function as estimated from the most recent blind H {\sc i} survey
data. We also re-examine the mass-metallicity relation of \cite{Tremonti-04}
and show that there are strong residuals in this relation as a function of
$G_{HI}/S$.

\section{Data}\label{sec:data}

We use the Max Planck Institute for Astrophysics / Johns Hopkins University 
(MPA/JHU) SDSS DR4
database\footnote{http://www.mpa-garching.mpg.de/SDSS/DR4/} as our parent
sample. The sample comprises $\sim 4\times 10^5$ objects that have been
spectroscopically confirmed as galaxies and have data publically available in
the SDSS DR4 \citep{Adelman-McCarthy-06}. Details can be found in
\citet{Kauffmann-03c} and \citet{Brinchmann-04}.
We also make use of data from the HyperLeda homogenized H {\sc i} catalogue
\citep{Paturel-03} and from the 2MASS all-sky extended source catalogue
\citep[XSC;][]{Jarrett-00}.

In order to make comparisons with K04, we first construct a sample of 721
galaxies ({\tt Sample I}) from cross-matching the parent SDSS sample, the
2MASS XSC and the HyperLeda H {\sc i} catalogue using the same selection
criteria as in K04. The galaxies are required to have positions matched to
HyperLeda objects within $6^{\prime\prime}$ (as in K04) and to 2MASS XSC
objects within $3^{\prime\prime}$ (as in \citet{Blanton-05}). Following K04,
the galaxies are also restricted to lie in the redshift range of $z<0.1$,
$r<17.77$ and have $K<15$, as well as reliable redshifts and magnitudes based
on data flags and errors (magnitude errors $<0.3$ in $K$, $<0.4$ in H {\sc i},
and $<0.15$ in $u$, $g$ and $r$). Here $u$, $g$ and $r$ are the SDSS Petrosian
apparent magnitudes, and $K$ is the 2MASS $K-$band extrapolated total
magnitude. 

Our second sample ({\tt Sample II}) consists of $800$ galaxies from the
cross-match of the SDSS DR4 and HyperLeda H {\sc i} data. The selection
criteria are the same as above, except that the 2MASS-related criteria are not
required here. In addition, we have visually examined the $r$-band image of
each galaxy and dropped those galaxies that have companion galaxies within
200 arcseconds\footnote{We have also selected a smaller sample consisting
of 129 galaxies that have no companions within 10 arcminutes and found no
significant change in the resulting H {\sc i} gas estimator.},
to avoid mis-estimatimg the H {\sc i} mass of the main galaxy.
This sample will be used to derive our best estimator
of the H {\sc i} mass fraction. We note that the median redshift of the
galaxies in both {\tt Sample I} and {\tt Sample II} is very low ($\sim 0.014$; 
see Figure~\ref{fig:sample_ii} below). 

Our third sample ({\tt Sample III}) is based on SDSS DR4. Because the
calibration sample ({\tt Sample II}) is limited at $z<0.1$, we apply the same
redshift cut to the DR4 sample. We also apply the same magnitude cuts to both
samples ($r<17.77$). Finally, a galaxy is only included in {\tt Sample III} if
there is a significant detection of the H$\alpha$ emission line in its optical
spectrum, that is, if $EW(H\alpha)> 3\sigma$. Here $EW(H\alpha)$ is the
equivalent width of the H$\alpha$ emission line and $\sigma$ is its error. Both
quantities are taken from the MPA/JHU database. This gives rise to a sample of
157,662 galaxies. We will use this sample to estimate the H {\sc i} mass
function and compare it to the results from real H {\sc i} surveys (\S~\ref{sec:himf}).
We note that more than 99\% of galaxies in Sample II show $EW(H\alpha)> 3\sigma$.

Finally we construct a fourth sample ({\tt Sample IV}), which is also based on
SDSS DR4, and consists of 64,305 star-forming galaxies with $r<17.77$ and
$z<0.1$, high S/N emission lines (S/N $>$ 3 for all the four emission lines on 
BPT (Baldwin-Phillips-Terlevich) diagram, see Brinchmann et al. 2004 for details.) 
and reliable estimates of oxygen
abundances (magnitude errors $<0.15$ in the five SDSS photometric bands;
metallicities in the range of $7.5<12+\log(O/H)<9.5$ ). We use this sample to
study whether the relation between stellar mass and gas-phase metallicity
depends on the amount of gas in the galaxy (\S~\ref{sec:mzr}).

\begin{figure*}
\centerline{\epsfig{figure=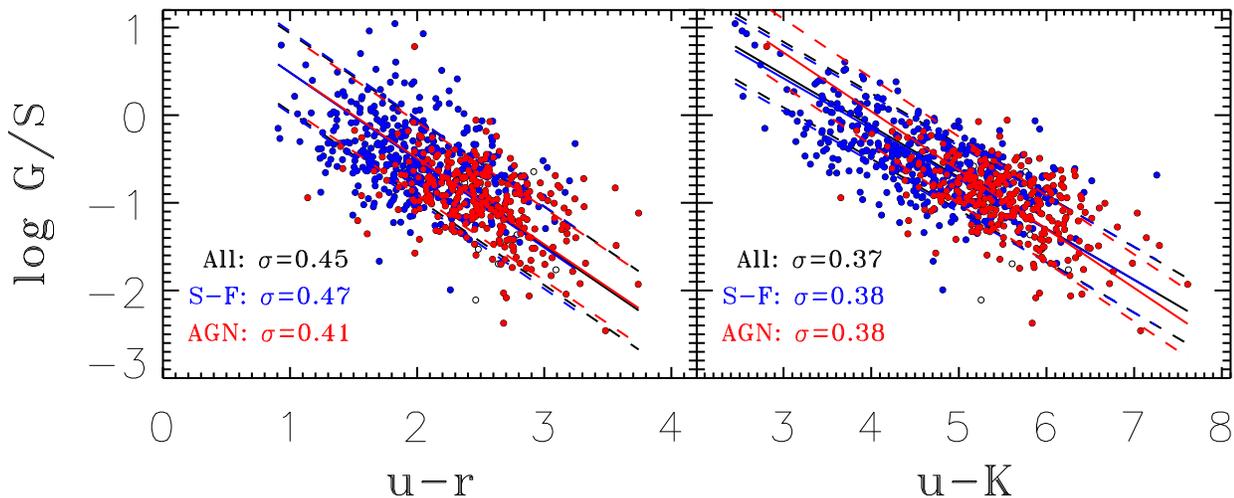,clip=true,width=\textwidth}}
\caption{Correlation of the atomic gas-to-stellar mass ratio, $G/S$, with
$u-r$ (left panel) and $u-K$ (right panel) colours, for 721 galaxies in {\tt
Sample I} that are selected from the main sample of SDSS DR4 and also have data from HyperLeda
and 2MASS. Galaxies are plotted in red dots if they are classified as AGN, blue dots
if they are classified as high S/N star-forming galaxies, and open circles if they 
are unclassified by \citet{Brinchmann-04}. Gas masses are derived 
from H {\sc i} fluxes with a helium correction factor of 1.4 as in 
\citet{Kannappan-04}, and stellar masses are derived from $K-$band fluxes using 
stellar mass-to-light ($M/L$) ratios estimated from $g-r$ colours as in 
\citet{Bell-03}. The three lines are the best-fitting linear relation 
(solid) and the $1\sigma$ variance (dashed), which are shown in red for the AGN,
in blue for the high S/N star-forming galaxies, and in black for the galaxies as 
a whole. The best-fit relations for the whole sample are $\log(G/S)=1.48-0.99(u-r)$ and
$\log(G/S)=2.19-0.58(u-K)$ with $\sigma=0.45$ and $0.37$ dex, in good agreement
with \citet{Kannappan-04} who found $\log(G/S)=1.46-1.06(u-r)$ and
$\log(G/S)=1.87-0.56(u-K)$ with $\sigma=0.42$ and $0.37$ dex.}
\label{fig:kannappan}
\end{figure*}

Throughout this paper we use stellar masses estimated from the $i$-band
luminosity and $g-r$ colour using the formula provided in B03, that is,
$\log(M_*/L_i)=-0.222+0.864(g-r)$.  The surface brightness used here is defined
as $\mu_i=m_i+2.5\log(2\pi R_{50}^2)$, where $m_i$ is the apparent Petrosian
$i$-band magnitude and $R_{50}$ is the radius (in units of arcsecond) enclosing
50\% of the total Petrosian $i$-band flux.  The stellar surface mass density is
given by $\log(\mu_*)=\log(M_\ast)-\log(2\pi R_{50}^2)$, where $M_\ast$ is the
stellar mass and $R_{50}$ is defined in the same way was as above but in units
of kpc.  The SDSS apparent magnitudes are corrected for foreground extinction
and are $k-$corrected to their $z=0$ value using the {\tt kcorrect v4.1.4} code
of \citet{Blanton-03b}.  The 2MASS $K-$band magnitude is $k-$corrected using
$k(z)=-2.1z$ (see B03).  Other paparameters such as star formation rate (SFR),
oxygen abundance, and emission line fluxes are taken from the MPA/JHU database.
The reader is referred to \citet{Brinchmann-04}, \citet{Kauffmann-03c} and
\citet{Tremonti-04} for detailed description of how these quantities are
derived. We use the {\em total} SFR for which the aperture bias is corrected
using resolved imaging.  Throughout this paper we assume a cosmological model
with $\Omega_0=0.3$, $\Lambda_0=0.7$, and $H_0=70$ kms$^{-1}$Mpc$^{-1}$.

\section{Estimating gas masses}\label{sec:estimators}

\subsection{Correlations of $G_{HI}/S$ with galaxy properties}

In Fig.~\ref{fig:kannappan} we plot the correlation of the atomic
gas-to-stellar mass ratio, $G/S$, with $u-r$ (left panel) and $u-K$ (right
panel) colours for the 721 galaxies in {\tt Sample I}. The atomic gas masses
and the stellar masses plotted here are estimated in exactly the same way as in
K04. Briefly, gas masses are derived from H {\sc i} fluxes with a helium
correction factor of 1.4, while stellar masses are from $K-$band fluxes using
stellar mass-to-light ($M/L$) ratios estimated from $g-r$ colours as in B03.
The only difference from the analysis of K04 is the fact that our sample is
based on a later SDSS data release. 
Using a sample of 346 galaxies constructed from cross-matching 
the SDSS DR2, the 2MASS XSC and the HyperLeda H {\sc i} catalogue, K04
found $\log(G/S)=1.46-1.06(u-r)$ and $\log(G/S)=1.87-0.56(u-K)$ with $1\sigma$
variance $\sigma=0.42$ and $0.37$ dex.
As can be seen from Figure~\ref{fig:kannappan}, the result
of K04 is well reproduced, both in the overall amplitude and slope of the mean
correlations, and in the scatter about the mean. We also divide our sample into
star-forming galaxies and AGN using the standard BPT classification diagram
\citep[see][for details]{Kauffmann-03c, Brinchmann-04} and plot these two
classes using different colour codings. As can be seen, AGN and star-forming
galaxies do not show significant difference in their correlation between gas
fraction and colour. We therefore do not remove AGN from any of our samples,
with the exception of sample IV, because we cannot obtain accurate gas-phase
metallicity measurements for AGN. 

\begin{figure*}
\centerline{\epsfig{figure=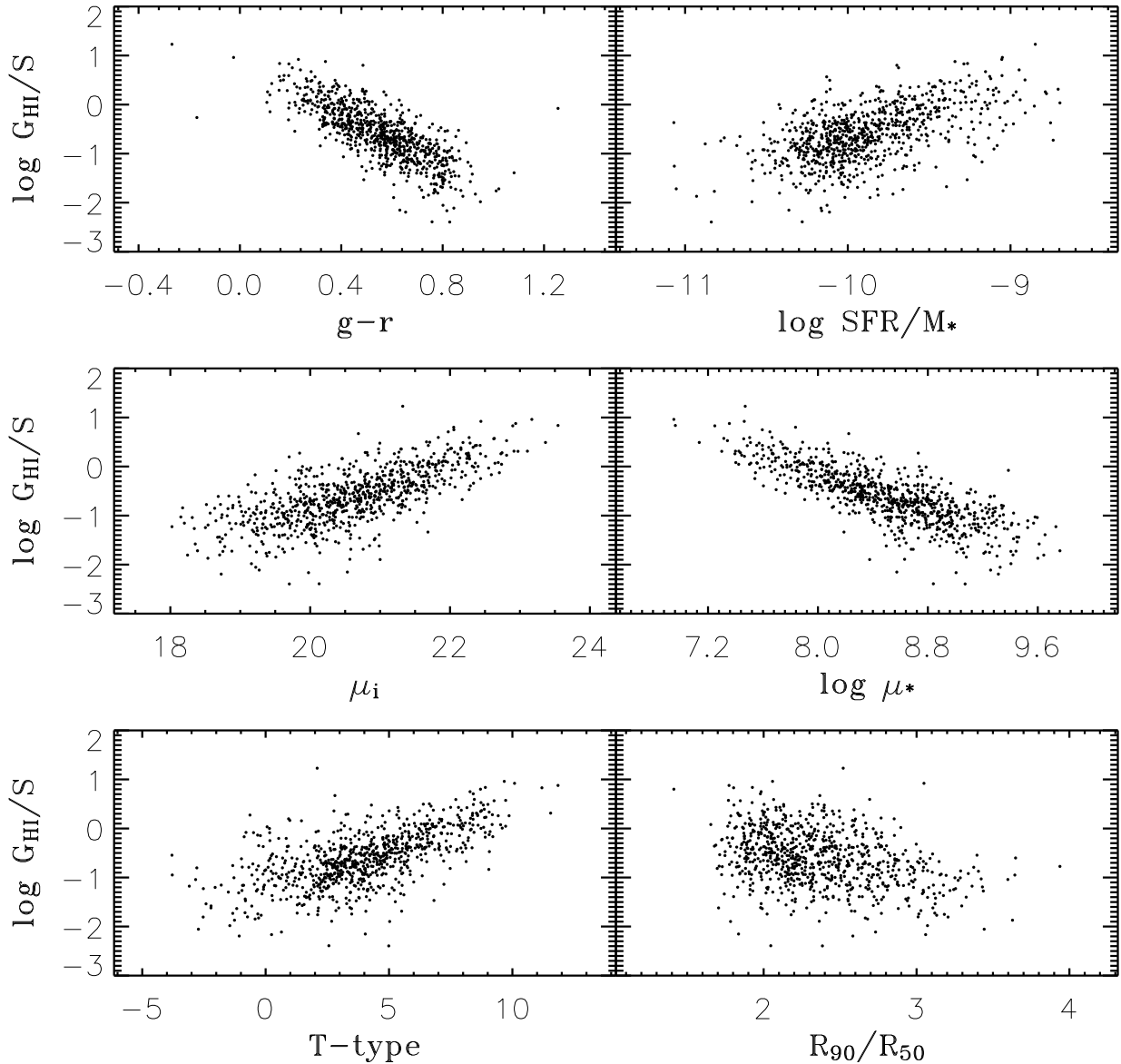,clip=true,width=\textwidth}}
\caption{Correlation of the atomic gas-to-stellar mass ratio, $G/S$, with
$g-r$ colour (top left), aperture corrected specific star formation rate from
\citet{Brinchmann-04} (top right), $i$-band surface brightness (middle left),
stellar surface mass density (middle right), T-type (bottom left, see text
for more explanation), and concentration index (bottom right) for 800 galaxies
in {\tt Sample II}}
\label{fig:correls}
\end{figure*}

We then examine the correlations between $G_{HI}/S$ and a variety of physical
quantities, in the hope of finding a even better estimator for the gas
fraction. It has long been established that the fraction of atomic gas
correlates with a variety of galaxy properties. It would certainly not be
surprising to find that a higher gas content is associated with galaxies with
lower stellar masses, bluer colours, lower surface brightnesses, disk-dominated
morphologies and spectral types indicative of the presence of a young stellar
population. In our calibrating sample, $G_{HI}/S$ exhibits the {\em tightest}
correlations with colour, surface brightness and stellar surface mass density,
with a typical scatter of $\sim 0.4$ dex (see Fig.~\ref{fig:correls} ).
Interestingly, we find that the correlation between $G_{HI}/S$ and $i$-band
surface brightness or surface stellar mass density is much tighter than the
correlation between $G_{HI}/S$ and concentration index, which provides a good
measure of the bulge-to-disk ratio of the galaxy \citep{Gadotti-09}. 

%We have also examined another indicator of morphology, the T-type from the RC3
%catalogue \citep{Vaucouleurs-91}. Out of the 800 galaxies in {\tt Sample
%II}, 527 have the morphological type, $T_{RC3}$, from this catalogue. For the
%remainder, this quantity is estimated using a back propagation neural
%network(BPNN). 
We have also examined another indicator of morphology, T-type \citep{Vaucouleurs-91}. 
We estimate this quantity using a back propagation neural network (BPNN).
The basic methodology we employ (number of layers and neurons,
the transfer function, the training algorithm etc) are exactly the same as
described in \cite{Ball-04}. The training and test samples are drawn from the RC3
and consist of 4393 galaxies with imaging data from the SDSS DR4. We follow
\cite{Fukugita-07} and use 29 photometric parameters, which are available
from SDSS, as the input for the BPNN. The inner structure of the network is
adjusted iteratively until an optimally trained network that links $T$ and
the available set of photometric parameters is obtained. The network is then
applied to the galaxies in our sample with no $T$ measurement. As can be seen
from the figure, the correlation of $G_{HI}/S$ and T-type is also weaker than
the correlation with surface mass density or surface brightness.

We find that $G_{HI}/S$ correlates very strongly with galaxy colour and
reasonably strongly with specific star formation rate (corrected for aperture
affects as described in \citealt{Brinchmann-04}). We also find correlations
with other quantities, but since all of them exhibit considerably more scatter,
we do not show them here.

\subsection{Deriving new $G_{HI}/S$ estimators}

\begin{figure*}
\centerline{\epsfig{figure=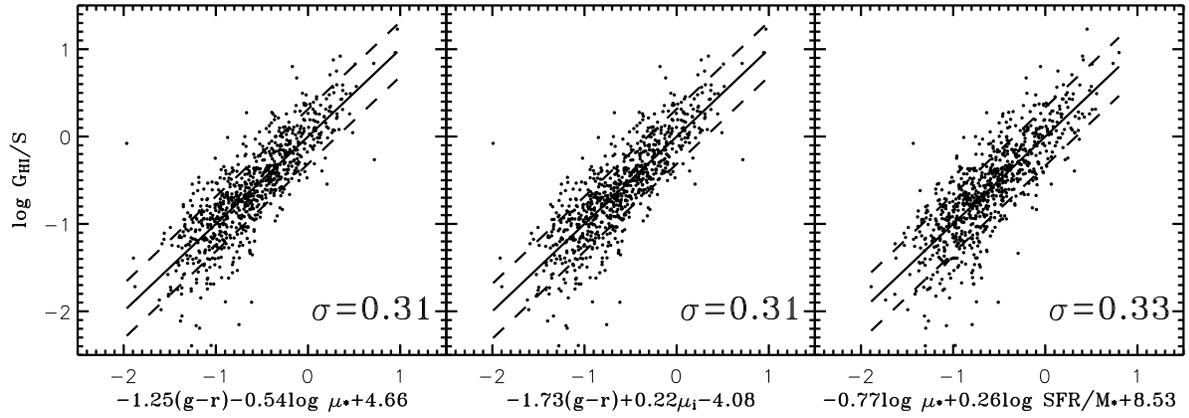,width=\textwidth,clip=true}}
\caption{We plot the relations between $G_{HI}/S$ and
the linear combinations of, a) $g-r$ colour and stellar surface
mass density (left), b) $g-r$ colour and $i$-band surface brightness
(middle), and c) aperture corrected specific star formation rate
from \citet{Brinchmann-04} and stellar surface mass density,
that minimize the scatter.}
\label{fig:estimator}
\end{figure*}

In this paper, we will only concern ourselves with our two strongest
correlations: i.e the correlation between  $G_{HI}/S$ and colour or specific star
formation rate and the correlation between  $G_{HI}/S$ and surface
brightness or stellar surface mass density. How can these correlations be understood?
Let us consider the  Kennicutt-Schmidt law of  star formation
\citep{Schmidt-63, Kennicutt-98} in  which the surface density of star
formation rate scales with the surface density of total (atomic+molecular) cold
gas as an increasing power law,
\begin{equation}
\Sigma_{SFR} \propto \Sigma_{gas}^n, 
\end{equation}
with a slope of $n\approx 1.4$. If the two surface densities are estimated
over the same radius, one can easily rewrite the Kennicutt-Schimit law as
follows: 
{\begin{equation}
SFR/M_\ast \propto (G/S)^n \mu_\ast^{n-1},
\end{equation}
where $G/S\equiv M_{gas}/M_\ast$, and $M_{gas}$ and $M_\ast$ are 
the total mass of cold gas and stars.} It is thus natural to expect
the gas mass-to-stellar mass ratio to correlate with these
properties, just as shown in Figure~\ref{fig:correls}. More
interestingly, the above equation implies that there exists a plane of
correlations, similar to the fundamental plane of early-type galaxies,
involving the following three variables: the gas-to-stellar mass ratio ($G/S$),
the specific star formation rate ($SFR/M_\ast$) and the surface stellar mass
density ($\mu_\ast$). This plane can be defined by 
\begin{equation}\label{eqn:spec_estimator}
 \log G/S = \log M_{gas}/M_\ast = a \log \mu_\ast + b \log SFR/M_\ast
 + c,
\end{equation}
where the coefficients $a$, $b$, and $c$ can be determined by minimizing the
residuals from the plane. If the gas content parameter $G/S$ that enters this
plane scales linearly with the observed H {\sc i}-to stellar mass ratio (we
admit that this may well be an over-simplification of the true situation), then
$ \log G_{HI}/S$ can also be expressed as a linear combination of the logarithm
of the stellar surface mass density and the logarithm of the specific star
formation rate.

Our best-fit relation between $G_{HI}/S$ and the linear combination of 
$\mu_\ast$ and $SFR/M_\ast$ (from {\tt Sample II}) is shown in the far-right panel
of Figure~\ref{fig:estimator}. The galaxies are plotted as black dots, and
the best-fit relation is shown as a solid line. The $1\sigma$ scatter
around the relation is 0.33 dex, which is significantly smaller than the
scatter in the relation between $ \log G_{HI}/S$ and stellar surface mass
density or the relation between $ \log G_{HI}/S$ and specific star formation
rate.

This relation is also tighter than the correlations of $G_{HI}/S$ with all
the other galaxy properties we have investigated. Thus, this relation can
serve as a better estimator of H {\sc i}-to-stellar mass ratio, compared to
those derived using a single parameter. 

We note that the correlation between colour and $G_{HI}/S$ shown in
Figure~\ref{fig:correls} is actually somewhat {\em tighter} than the
correlation between SFR/$M_*$ and $G_{HI}/S$. One reason for this may be that the
Brinchmann et al (2004) aperture corrections are not accurate. 
As shown in \citet{Brinchmann-04} (see their Fig.~14), for 
$\log(SFR/M_\ast)>-10.5$, the 2$\sigma$ uncertainties on the aperture-corrected 
$\log$ SFRs are larger by $\sim0.3$ dex than for $\log$ SFR measured inside the 
fibre because the aperture corrections are significantly more uncertain than 
the SFR estimates from the spectra.
The errors will be even larger for the galaxies in our sample, because 
the median redshift is much lower and the total star formation rates are
derived almost entirely from the photometry, not from the emission lines
measured in the spectra.

In addition, we
find the correlation between $G_{HI}/S$ and surface brightness is just as tight
as the correlation between $G_{HI}/S$ and stellar surface mass density, so
there is no real value in working with the \citet{Brinchmann-04}
star formation rate estimates  , rather than with directly measured colours.
For our purposes, this is in fact quite encouraging,
because it suggests that one can get a reasonably good prediction for the
fraction of atomic gas in a galaxy from photometrically measured quantities.
This is most frequently available for high redshift galaxies.

We propose that the ($g-r$) colour and $i$-band surface brightness $\mu_i$
are good choices for this purpose. The correlation plane that uses the
quantities ($g-r$) and $\mu_\ast$ is shown in the leftmost panel in
Figure~\ref{fig:estimator}, and the plane that uses the quantities ($g-r$) and
$\mu_i$ is shown in the central panel. The scatters in the relations
are the same (0.31 dex), even smaller than the plane involving $SFR/M_\ast$
and $\mu_\ast$. We thus adopt the relation in the central panel, 
\begin{equation}\label{eqn:estimator}
\log_{10}(G_{HI}/S) = -1.73238(g-r) + 0.215182\mu_i - 4.08451,
\end{equation}
as our final estimator of $G_{HI}/S$. The scatter in our new estimator represents
a 20\% decrease as compared in that of K04.

\begin{figure*}
\centerline{\epsfig{figure=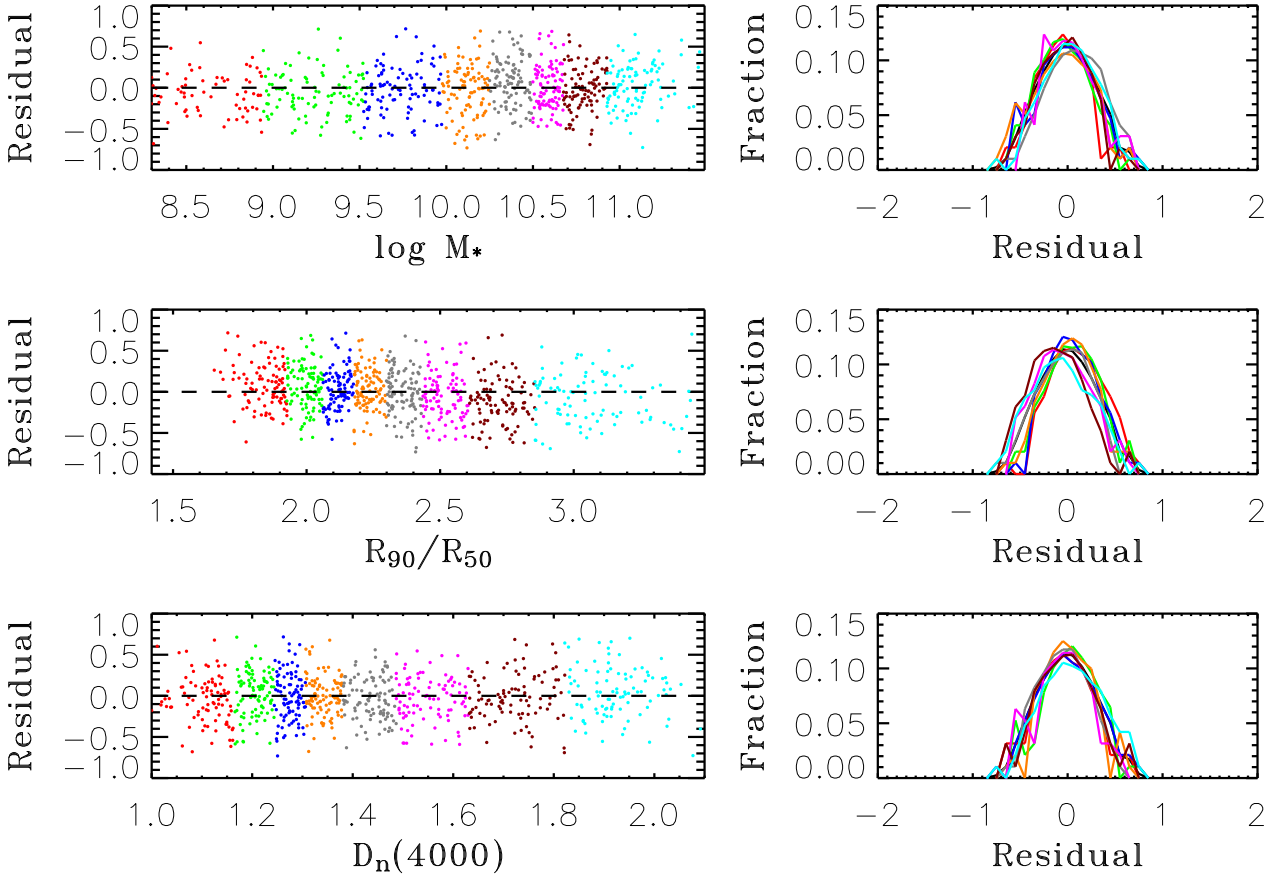,width=\textwidth,clip=true}}
\caption{We plot the residual in the predicted value of $G_{HI}/S$ from equation (4)
as a function of stellar mass (top left), concentration index (middle left)
and 4000 \AA\ break strength (bottom left). We have also divided the
galaxies into 8 subsamples as indicated by the coloured dots (left panels) or lines (right
panels). The left panels show the residual for individual galaxies, while the right panels 
show the distribution of the residuals in each of the subsamples.}
\label{fig:residuls}
\end{figure*}

\begin{figure*}
\centerline{\epsfig{figure=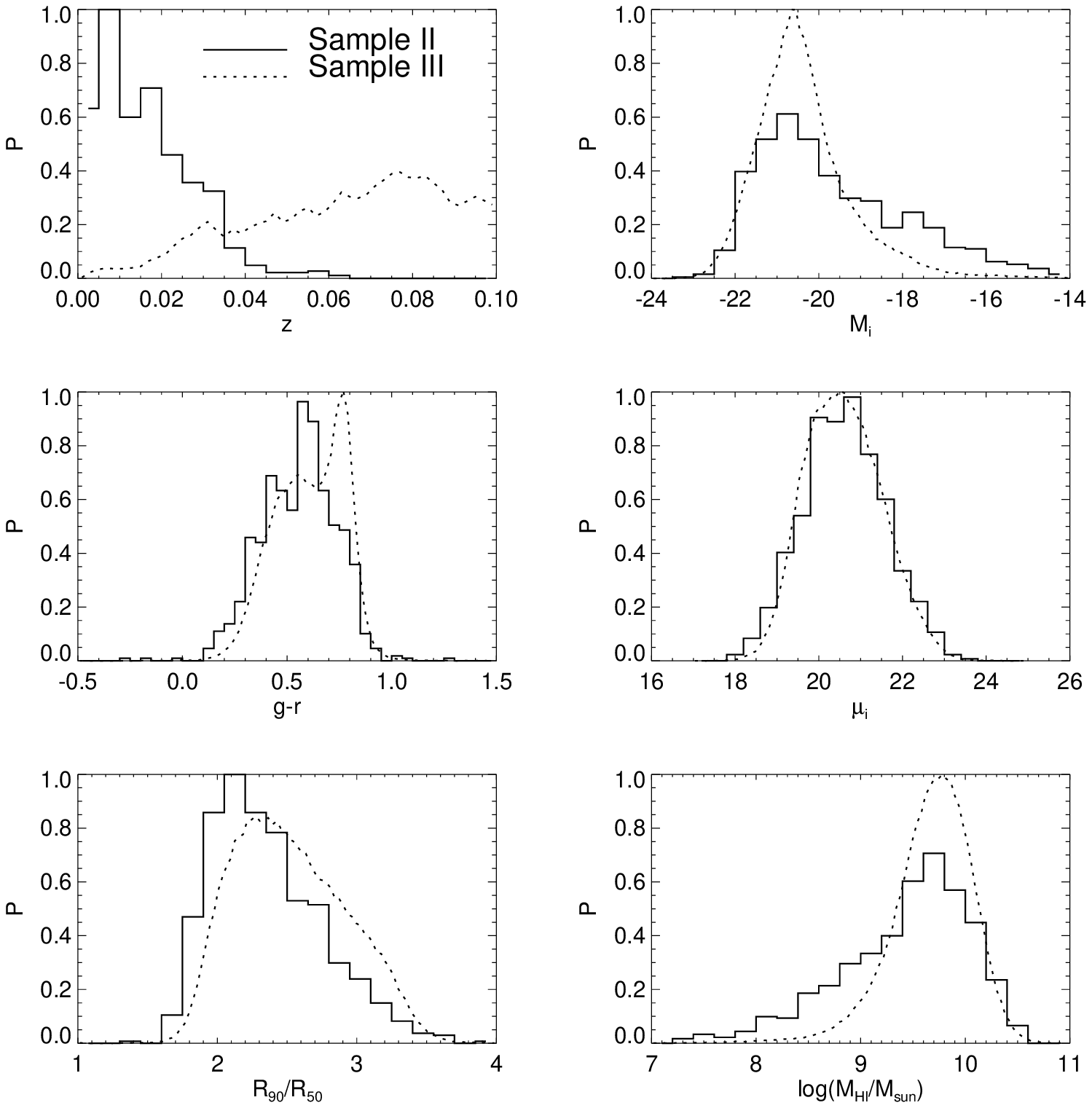,width=\textwidth,clip=true}}
\caption{Distributions of redshift, $i$-band absolute magnitude, $g-r$ colour, 
$i$-band surface brightness, concentration parameter $R_{90}/R_{50}$ measured in $i$-band,
and H {\sc i} gas mass, for galaxies in our calibration sample ({\tt Sample II}, solid
line) and in {\tt Sample III} (dotted line). In each panel the maximum of the two curves
is set to unity and the two curves are normalized so that they enclose the same area.
}
\label{fig:sample_ii}
\end{figure*}

Before we go ahead and apply this estimator to large samples or try to
extrapalate it to higher redshifts, it is important to test whether we can find
any systematic effects that could bias such analyses. For example, we know that
pure passive aging of stellar populations will cause the colour of a galaxy to
evolve with time. It would therefore not be surprising if $G_{HI}/S$ at a fixed
value of the $g-r$ colour was to depend on redshift. In the absence of H {\sc i} data
for high redshift galaxies, the only way we can test for such effects is to see
whether the {\em residuals} around our best-fit estimator are correlated with
intrinsic properties of the galaxies, such as stellar mass, morphology or mean
stellar age. In Figure~\ref{fig:residuls} we plot the residuals of $G_{HI}/S$
from the relation given in Eq.(\ref{eqn:estimator}) as a function of stellar
mass, concentration index and 4000 \AA\ break strength. There is no clear
tendency for the $G_{HI}/S$ residuals to correlate with stellar mass or with
4000 \AA\ break strength. There is a small, but significant trend in the
residual as a function of the concentration index, in the sense that our
estimator overpredicts the gas fraction for the least concentrated galaxies and
underpredicts for the most concentrated objects. The effect is not large ---
$\sim 0.2$ dex shift in the predicted value of $\log G_{HI}/S$ from the least
concentrated to the most concentrated galaxies in our sample. We do not see
significant trends in the scatter in the residuals as a function of any galaxy
property, suggesting that it may be possible to calibrate out such systematic
trends in the future. 

We caution that our calibrating sample is compiled from many different samples 
with different selection effects, and may thus be biased because it is {\em not} 
a truly representative sample of nearby galaxies. This can be seen from 
Figure~\ref{fig:sample_ii} where we plot the histograms of a variety of physical
properties for galaxies in {\tt Sample II}.
It will be important to re-examine these issues with larger and more homogeneous
galaxy samples, as will be provided by the ALFALFA survey. 

\begin{figure*}
\centerline{
\epsfig{figure=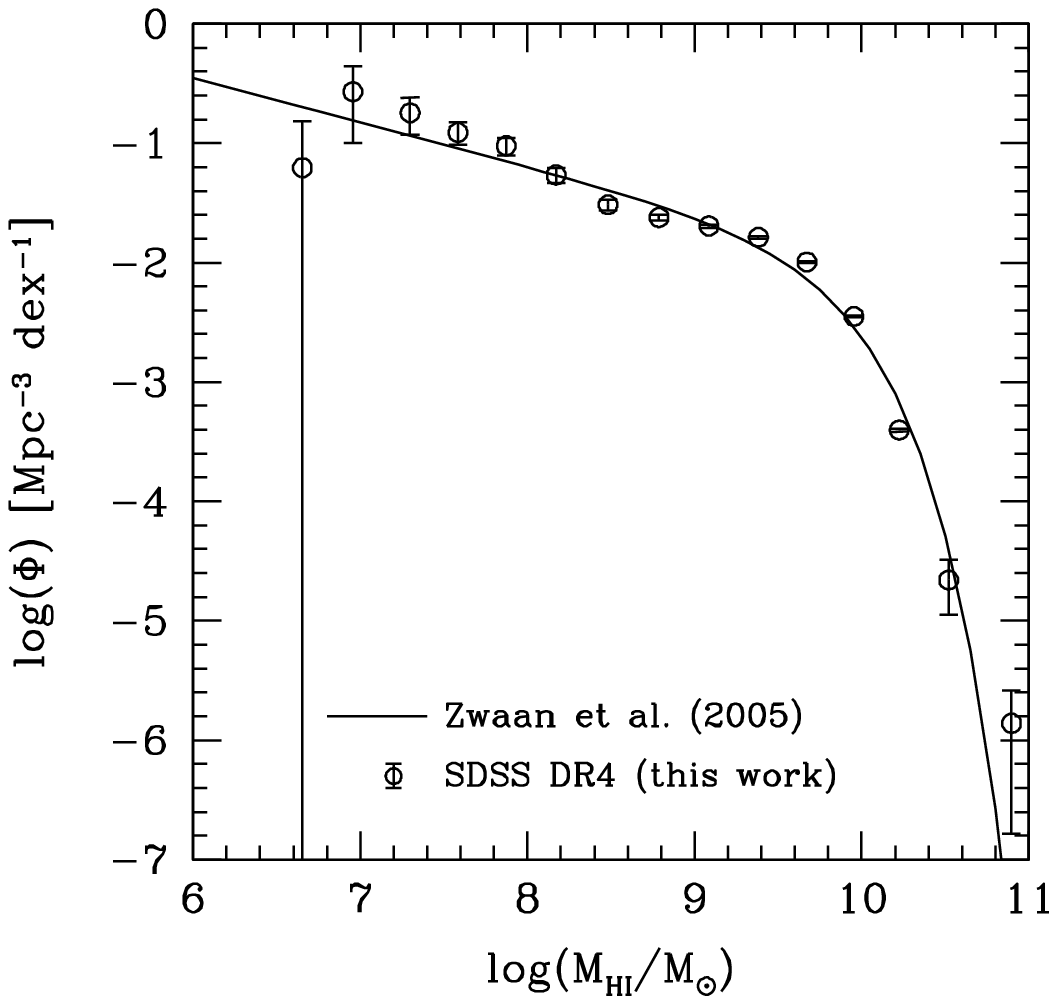,clip=true,width=0.33\textwidth}
\epsfig{figure=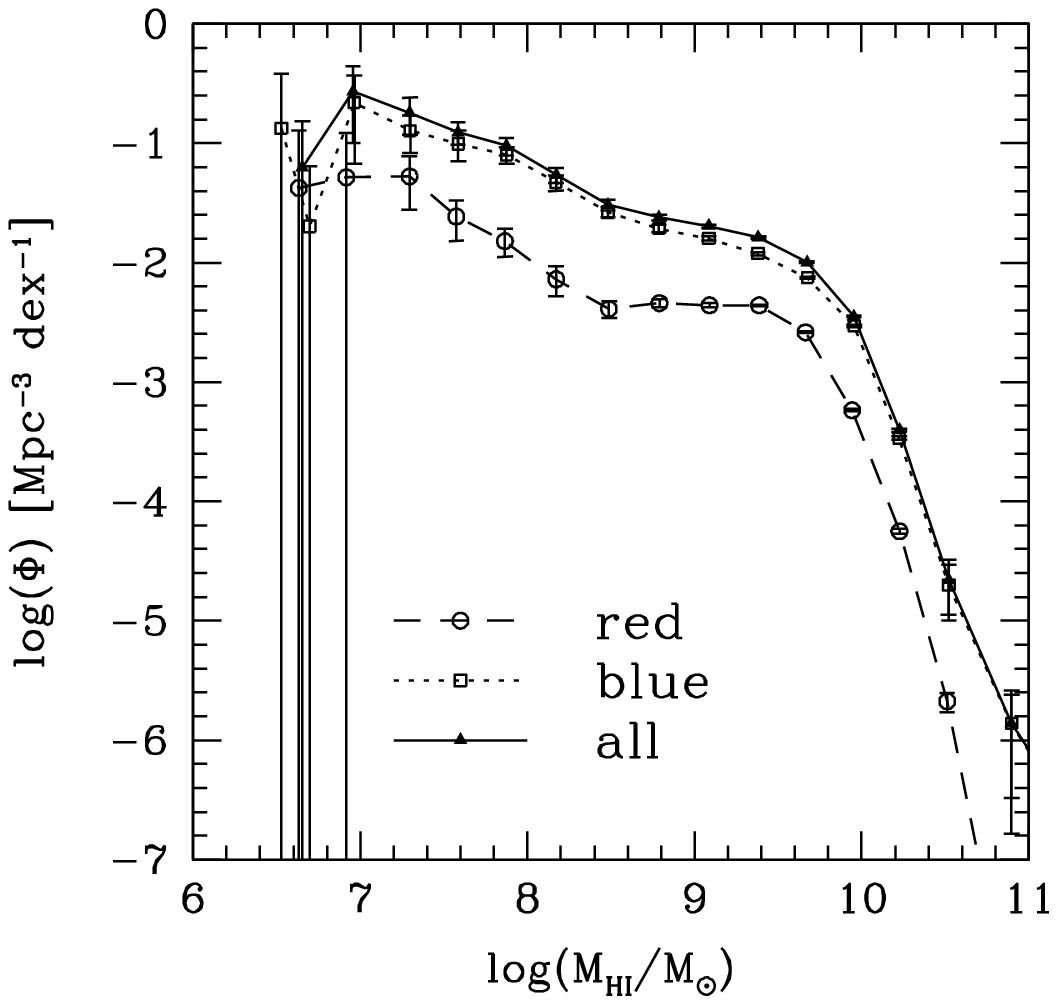,clip=true,width=0.33\textwidth}
\epsfig{figure=f6c.ps,clip=true,width=0.33\textwidth}
}
\caption{{\it Left:} H {\sc i} mass function for the galaxies in the
 SDSS DR4 that have redshift below 0.1 and evident H$\alpha$ emission
 in their optical spectrum is plotted as symbols, compared to the
 result from previous study which is based on real H {\sc i}
 observation and is plotted as line. {\it Center:} H {\sc
 i} mass function is plotted for red/blue galaxies separately. The
 result for the full sample is repeated from the left panel.
{\it Right:} redshift distributions for red and blue galaxies 
as well as for the galaxies as a whole are plotted in the upper part. 
The lower part shows the redshift distribution for galaxies in the
full sample, but at three fixed stellar masses as indicated.}

\label{fig:himf}
\end{figure*}

\section{Applications}

We now illustrate two different applications of our $G_{H {\sc I}}/S$
indicator. We first use the colours and surface brightnesses
of the galaxies in Sample {\tt III} to estimate the 
 H {\sc i} mass function and we compare our estimate with recent
results from real H {\sc i} surveys. We then examine whether
residuals in the relation between gas phase metallicity and stellar
mass are correlated with the H {\sc i} content of galaxies.

\subsection{H {\sc i} Mass Function}\label{sec:himf}

For each galaxy $i$ in {\sc Sample III} we compute the quantity $z_{max,i}$,
which is defined as the maximum redshift at which the galaxy would satisfy the
apparent magnitude limit of the sample. k-corrections are included when
calculating $z_{max,i}$ as described in \cite{Blanton-03b} 
\citep[see also][]{Blanton-Roweis-07}.  We do not apply
evolutionary corrections, because our sample is constrained to lie at $z<0.1$.
$V_{max,i}$ is defined for each galaxy as the comoving volume of the survey out
to redshift $z_{max,i}$, or 0.1 if $z_{max,i}>0.1$. The H {\sc i} mass function 
is thus estimated as
\begin{equation}
\Phi(M_{HI})\Delta M_{HI} =
\sum_{i}\left(f_{sp,i}V_{max,i}\right)^{-1},
\end{equation}
where $f_{sb,i}$ is the spectroscopic completeness of the survey area where the
galaxy $i$ is located and is defined as the fraction of the photometrically
targeted galaxies in the area for which usable spectra were obtained. The sum
in the above equation extends over all galaxies with H {\sc i} mass in the
range $M_{HI}\pm 0.5\Delta M_{HI}$.

Figure~\ref{fig:himf} shows the H {\sc i} mass function for galaxies in {\tt
Sample III}. The error bars are estimated using the bootstrap resampling
technique \citep{Barrow-Bhavsar-Sonoda-84}. We generated 100 bootstrap samples
from {\tt Sample III} and computed the H {\sc i} mass function for each sample.
The errors are then given by the scatter of the mass function among these
samples. We would like to point out that these errors are underestimated,
because they account only for the uncertainties due to sampling variance, but
do not include the effect of cosmic variance, systematic uncertainties in the
estimates of stellar masses, and scatter in the estimates of H {\sc i} mass.
For comparison, we also plot in Figure~\ref{fig:himf} a recent measurement of
the H {\sc i} mass function by \cite{Zwaan-05}, based on a complete H {\sc
i}-selected sample of 4315 galaxies from the HI Parkes All Sky Survey (HIPASS). HIPASS achieves
100\% coverage over the southern sky and the samples used for calculating the H
{\sc i} mass functions lie in the redshift range $0.003 < z < 0.02$. Most
other determinations of the H {\sc i} mass function have been based on
significantly smaller samples \citep{Zwaan-97, Henning-00,
Rosenberg-Schneider-02, Zwaan-03}, or have been based on samples 
containing only specific morphological types \citep{Springob-Haynes-Giovanelli-05}.

Our derived H {\sc i} mass function is in excellent agreement with the \cite{Zwaan-05}
result and is reasonably well fit by a single Schechter function with
parameters close to those given by \citet{Zwaan-05}.
A comparison of our measured H {\sc i} mass function with their best-fit
Schechter function gives a reduced $\chi^2$ statistic of $\chi^2/d.o.f=24/12$.
Their parameters yield
slightly fewer galaxies at the low-mass end. A straightforward integration of
our H {\sc i} mass function gives the mean comoving H {\sc i} mass density of
the low-redshift Universe of $\rho_{HI}\sim 7.5\times 10^7 hM_\odot/$Mpc$^3$,
corresponding to a cosmological H {\sc i} mass density of $\Omega_{HI}\sim
2.7\times 10^{-4} h^{-1}$, again well consistent with \citet{Zwaan-05}. In the
standard concordance cosmology \citep[e.g.][]{Komatsu-09} this matter density
corresponds to only $\sim 0.8\%$ of the baryons in the low-redshift Universe in
H {\sc i} gas in galaxies, compared to $3.5\%$ in stars \citep{Li-White-09}.

It is also interesting to compare the relative abundance of red and blue
galaxies at given H {\sc i} mass. We split the sample into two colour bins
using the following, luminosity-dependent cut:
\begin{equation}
 (g-r) = -0.104-0.042 M_r.
\end{equation}
We then compute the H {\sc i} mass functions for the red and blue subsamples
separately and plot them in the right-hand panel of Figure~\ref{fig:himf}. The
result for the full sample is also plotted.  As can be seen, the H {\sc i} mass
function is dominated by blue galaxies at all masses.

\begin{figure*}
\centerline{
  \epsfig{figure=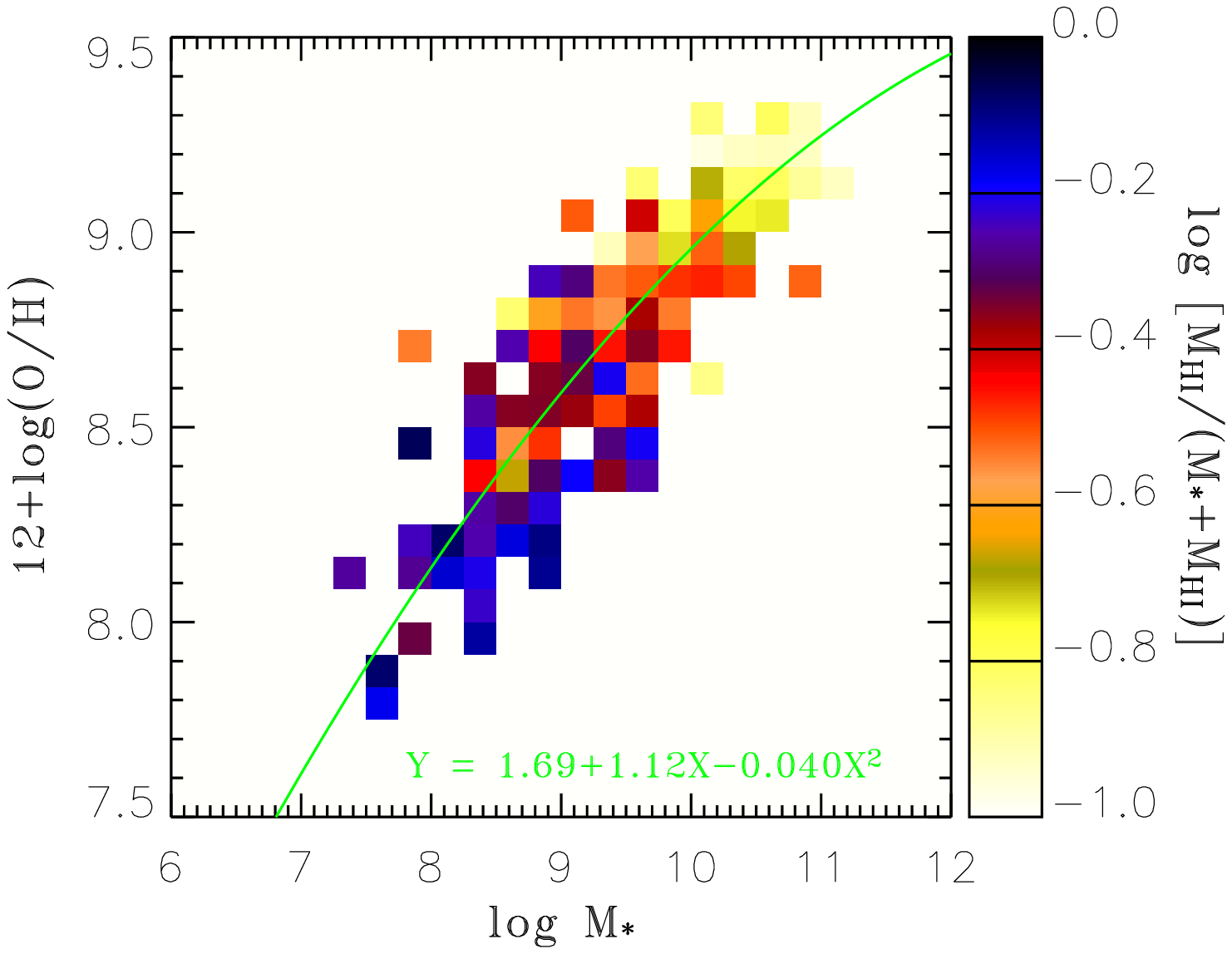,width=0.35\textheight,clip=true}
  \epsfig{figure=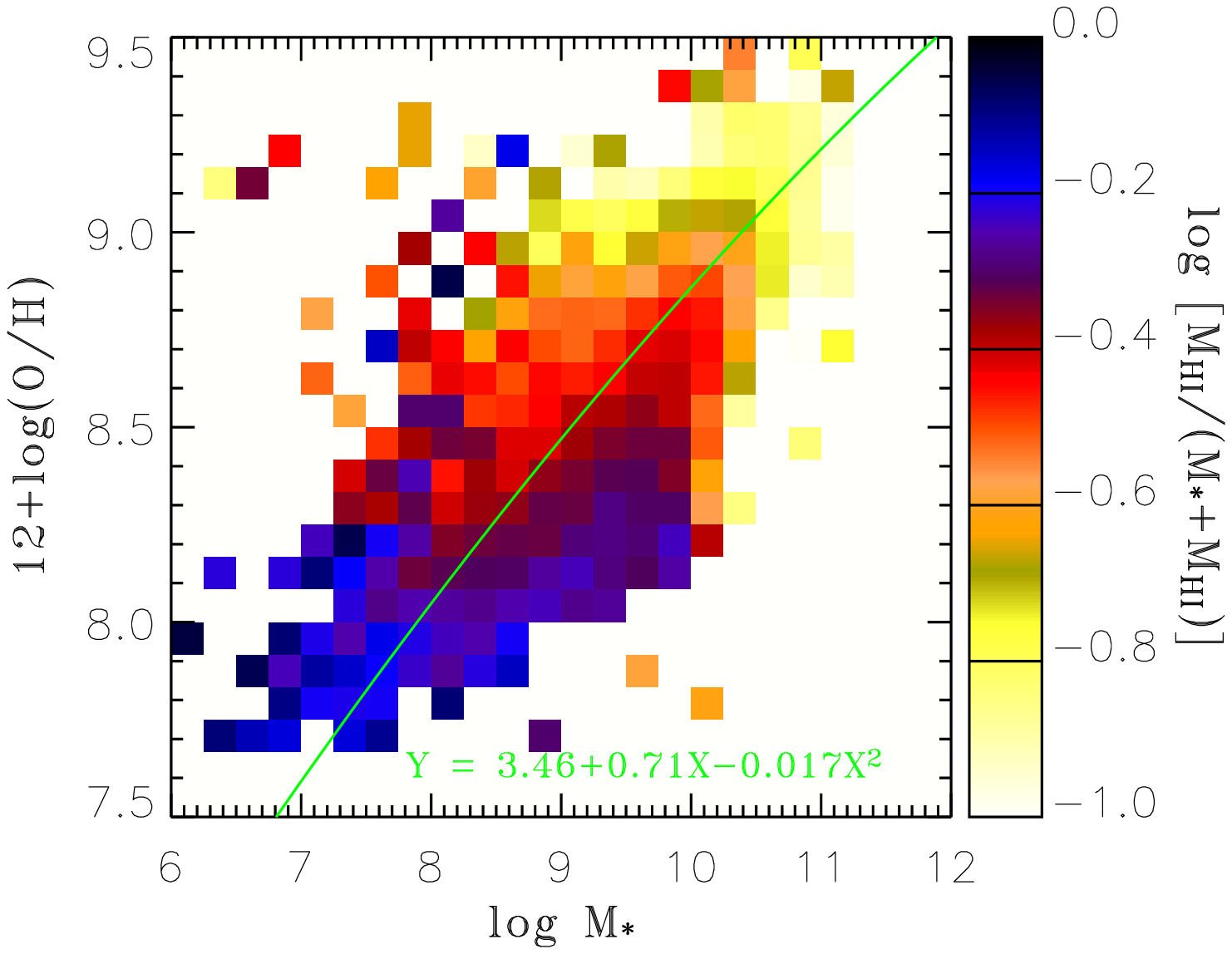,width=0.35\textheight,clip=true} }
\centerline{
  \epsfig{figure=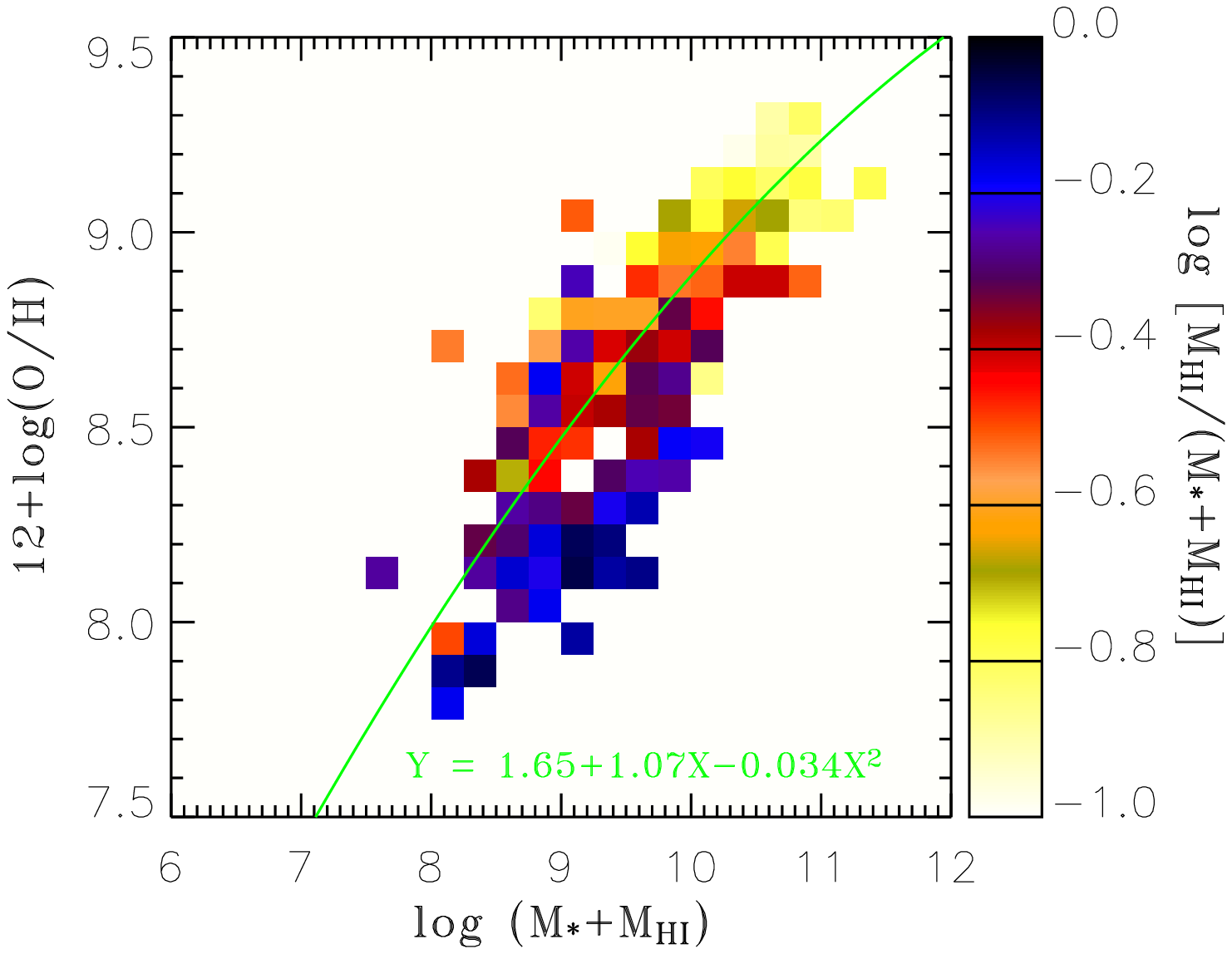,width=0.35\textheight,clip=true}
  \epsfig{figure=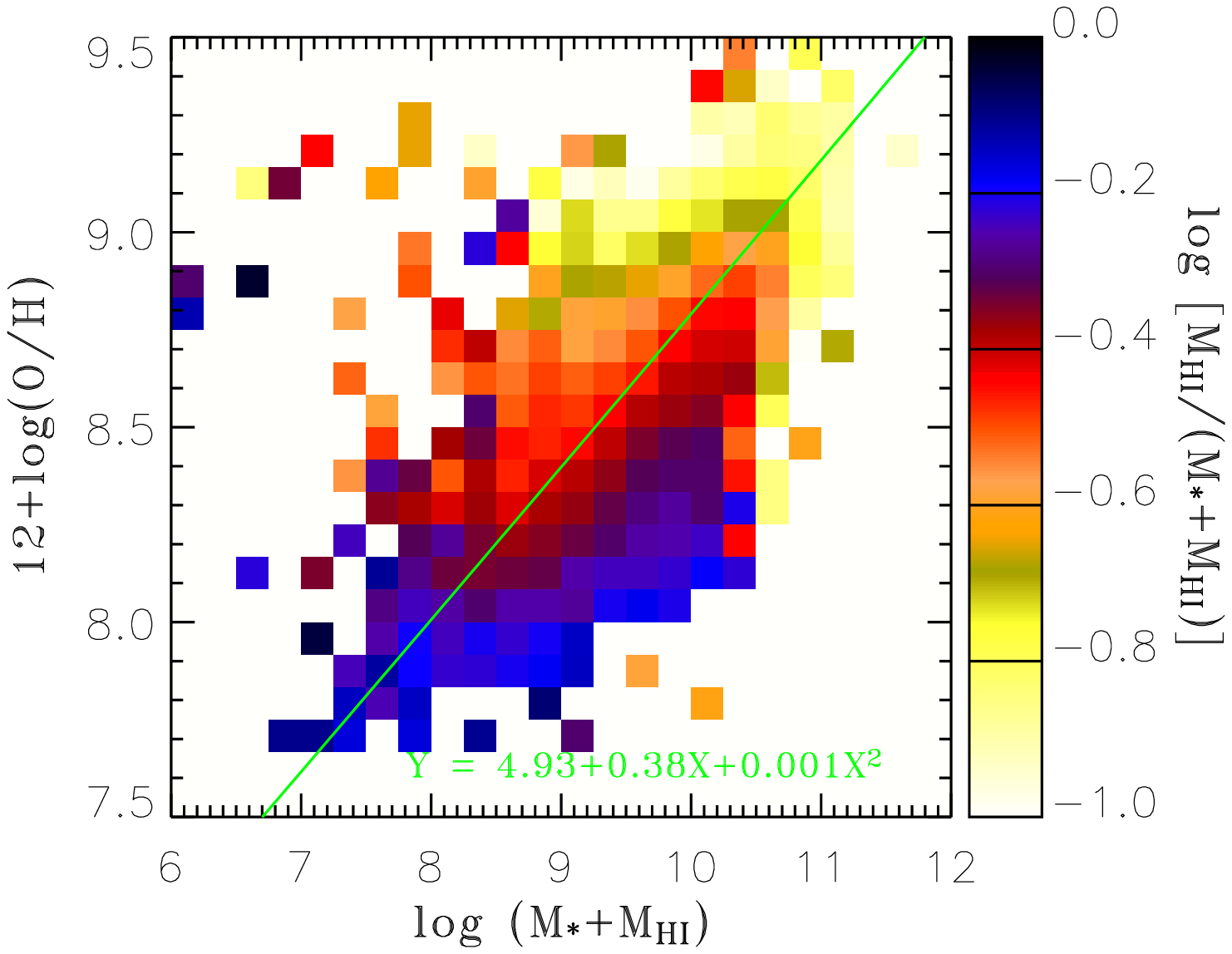,width=0.35\textheight,clip=true} }
\caption{Correlation of metallicity with stellar mass (top) and with
 stellar mass plus H {\sc i} gas mass (bottom).
 The left-hand panels show results for a
 sample of 800 galaxies from the HyperLeda catalogue
 for which the H {\sc i} mass is actually measured. The right hand panels show 
 results for $1.4\times 10^5$ star-forming galaxies in
 SDSS DR4 for which the H {\sc i} mass is given by our estimator
 based on the $i$-band surface brighteness and $g-r$ colour.
 Each pixel in the mass-metallicity plane has been colour-coded according to
 the mean H {\sc i} gas fraction of the galaxies that fall in that region. The
 green line in each panel shows the 2nd-order polynomial
 function that provides the best fit to the mean relation.}
\label{fig:mzr}
\end{figure*}

The redshift distribution of the red/blue subsamples as well as that of the
full sample is shown in the far-right panel in Figure~\ref{fig:himf}. 
The distribution of other physical properties of the full sample is plotted
in dashed lines in Figure~\ref{fig:sample_ii}.
The redshift distributions show strong features which reflect large-scale
structure. The structure at $z\sim 0.08$ is the well-known super cluster in SDSS,
the Sloan Great Wall. Two other weeker structures are also seen at redshifts
around 0.01 and 0.03. These structures are likely the reason why our H {\sc i} mass
function shows slightly higher (but still within error bars) amplitude than
that of Zwaan et al. (2005), at masses below $\sim10^9 M_\odot$ and at those
around the characteristic mass ($10^{9.8}M_\odot$). However, given the large
uncertainties in our H {\sc i} mass estimates and the small sample size of
Zwaan et al. (2005), such discrepancies should not be overemphasised.

%{\xtwfn The redshift distributions show strong features which reflect large-scale
%structure. The structure at z$\sim$0.08 is the well-known super cluster in SDSS,
%the Sloan Great Wall. The structure at z$\sim$0.03 is weaker but is mostly likely
%the reason why our H {\sc i} mass function show slightly higher (but still within
%error bars) amplitude at masses below 10$^8$ M$_\odot$, when compared to that of 
%\cite{ Zwaan-05}. These authors used a different method of estimating their H {\sc i} 
%mass function in which the effects of large-scale structure were carefully corrected. 
%The $1/V_{max}$-based method does not correct for these effects and so the resulting 
%H {\sc i} mass function is very sensitive to the structures at low redshifts as it is 
%dominantly contributed by these redshifts.}

\subsection{The Mass-Metallicity Relation}\label{sec:mzr}

The relationship between mass and metallicity \citep[MZR;][]{Lequeux-79,
Tremonti-04} is of particular interest in studies of galaxy formation and
evolution. There is now observational \citep[e.g.][]{Garnett-02, Tremonti-04}
evidence that metal loss via galactic winds \citep{Larson-74} may be largely
responsible for driving the relation. However, other processes may also be
important. For example, low-mass galaxies have higher gas fractions \citep[e.g.][]{Boselli-02}
and are hence expected to be less
enriched even if they were to evolve as closed boxes. It has even been proposed
that a variable IMF could also produce a mass-metallicity relation
\citep{Koppen-Weidner-Kroupa-07}. 

\begin{figure*}
\centerline{
 \epsfig{figure=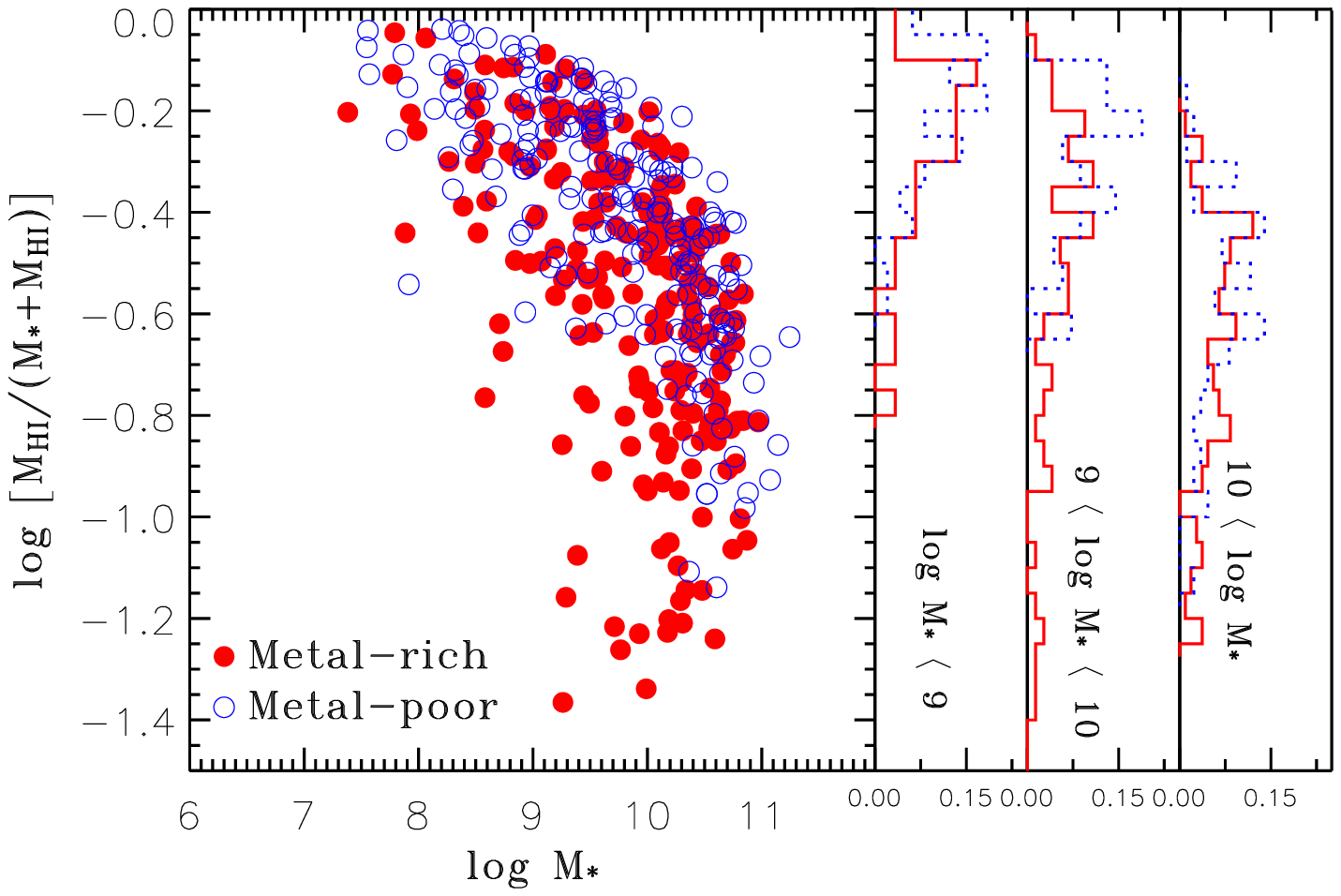,height=0.4\textwidth,clip=true}
 \epsfig{figure=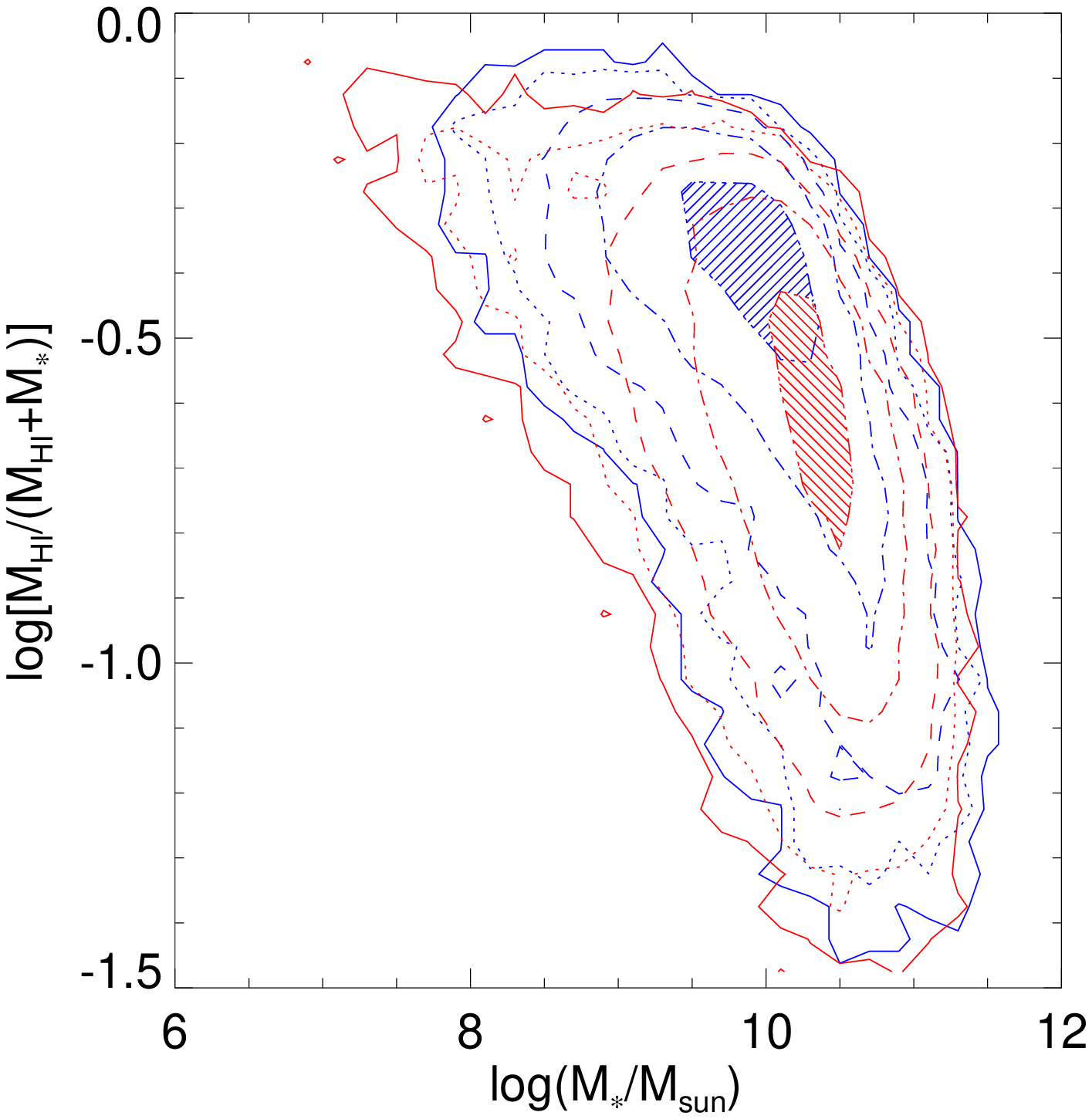,height=0.4\textwidth,clip=true} }
\caption{{\it Left:} H {\sc i} gas mass fraction versus stellar mass 
  relation for metal-rich (red) and metal-poor (blue) galaxies, for a
  small sample of galaxies that have real H {\sc i} observations.
  The three small panels compare the histogram of H {\sc i} gas fraction for metal-rich 
  (red solid) and metal-poor (blue dashed) galaxies in three stellar mass intervals as indicated.
  {\it Right:} contours of number density of galaxies in the plane of H {\sc i} gas
  mass fraction versus stellar mass, for a large sample of star-forming galaxies
  from the SDSS DR4 with H {\sc i} mass estimated from photometry. 
  Red and blue lines are for metal-rich and metal-poor galaxies respectively.
  The contour levels, which indicate the fraction of galaxies
  in the two subsamples that fall within a given region of the plane,
  are decreased by factors of 4 from the highest 
  (0.016 [0.2 $\log_{10}M_\odot$]$^{-1}$ [0.05 dex]$^{-1}$) to the lowest
  (6.25$\times10^{-5}$ [0.2 $\log_{10}M_\odot$]$^{-1}$ [0.05 dex]$^{-1}$).
  The region enclosed by the highest-level contour is shaded in red (blue) for the
  metal-rich (-poor) population.}
\label{fig:fgas_mstar}
\end{figure*}

In addition, the mass-metallicity relation has been found to depend on other
properties of the galaxies in the sample (for example surface mass density
[\citealt{Tremonti-04}], star formation rate as measured by ultraviolet-luminosity
and surface brightness [\citealt{Hoopes-07}], specific star formation rate and size
[\citealt{Ellison-08b}], the presence or absence of close companions
[e.g. \citealt{Michel-Dansac-08}], cluster membership and local density 
[\citealt{Ellison-09}] and also on large-scale environment
[e.g. \citealt{Cooper-08}]).  It would be useful to figure out which of these
dependencies are primary causes of variation in the relation, and which are
secondary effects. For example, if the mass-metallicity relation depends on
specific star formation rate, it will naturally also depend on 
environment, simply because the specific star formation rates of galaxies
depend strongly on density.

The gas mass fraction is the natural parameter for quantifying the degree to
which a galaxy has exhausted its available fuel supply and one would expect the
mass-metallicity relation to depend quite strongly on this quantity. In
Figure~\ref{fig:mzr}, we examine the trend of gas mass fraction, defined as the
ratio of H {\sc i} gas mass to the sum of H {\sc i} and stellar mass, with the location of
galaxies in the plane of metallicity versus stellar mass (top) and metallicity
versus H {\sc i} mass plus stellar mass (bottom). In each case, we compare
results for our small calibrating sample with real H {\sc i} masses ({\tt
Sample II}, left-hand panels) with results obtained for all high S/N
star-forming galaxies ({\tt Sample IV}, right-hand panels). As can be seen, the
basic trend is very similar for the two samples. Star-forming galaxies
with lower masses and metallicities have higher gas fractions.

In addition, Figure~\ref{fig:mzr} shows that at fixed stellar mass, metal poor
galaxies have higher H {\sc i} fractions than metal-rich galaxies. This is
clearly seen in Figure~\ref{fig:fgas_mstar} where we plot the gas
fraction---stellar mass relation for galaxies in {\tt Sample II} and {\tt
Sample IV}. Metal-poor galaxies are plotted in blue, while metal-rich galaxies
are plotted in red. In order to split our sample into two subsamples in
metallicity, we determine a stellar mass-dependent cut in metallicity by
fitting a 2nd order polynomial function to the stellar mass-metallicity
relation shown in the top two panels of Figure~\ref{fig:mzr}. A galaxy located
above (below) this cut is thus classified as metal-rich (-poor). As can be seen
from Figure~\ref{fig:fgas_mstar}, metal-poor galaxies are shifted almost
vertically in this diagram, towards higher H {\sc i} fractions. This is true
for both our calibrating sample and for the full sample of star-forming
galaxies. One possible interpretation of this result, is that the scatter of
metallicity at fixed stellar mass might be partially (if not totally) due to
recent inflow of less enriched gas from the surrounding halo. 

In order to test whether this explanation is plausible, one requires a model
for the chemical enrichment of galaxies that takes into account the effect of
supernova-driven winds and the infall of gas from the surrounding halo. Such
models already exist. For example, \cite{Lucia-Kauffmann-White-04} implemented
a model for the chemical enrichment of galaxies in a high resolution simulation
of a $\Lambda$CDM universe. The transport of metals between the stars, the cold
gas in galaxies, the hot gas in dark matter haloes and the intergalactic gas
outside virialized haloes was modelled in detail. In the scheme adopted by the
authors, metals are ejected outside the halo by supernovae and later
reincorporated when structure collapses on larger scales. The model also
followed the continued infall of cold gas onto the galaxy through merging of
gas-rich satellites and through cooling from a surrounding hot gas halo. After
suitable adjustments to the free parameters in the model, a good fit to the
observed relations between stellar mass, gas mass and metallicity was obtained. 

\begin{figure}
\centerline{\epsfig{figure=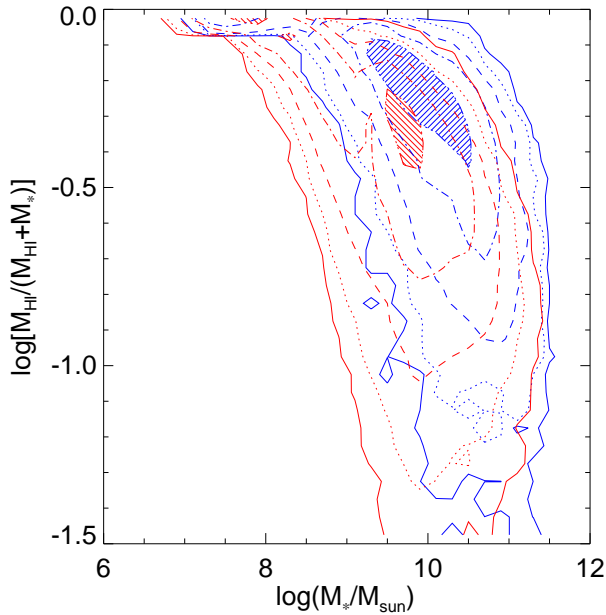,width=0.5\textwidth,clip=true}}
\caption{Gas mass fraction versus stellar mass relation for metal-rich (red)
and metal-poor (blue) galaxies in the $z=0$ semi-analytic catalogue of
\citet{Lucia-Blaizot-07}. The meaning of the colour coding and contour levels
is the same as in the right-hand panel of the previous figure.}
\label{fig:sam}
\end{figure}

We now test whether these same models can reproduce the secondary trend with
gas fraction that we see in our data. In Figure~\ref{fig:sam}, we plot the
stellar mass-gas fraction relation predicted by the semi-analytic model of of
\cite{Lucia-Blaizot-07}, which incorporates the same chemical enrichment scheme
as in \cite{Lucia-Kauffmann-White-04}. We have selected all galaxies in the z=0
catalogue (available at http://www.mpa-garching.mpg.de/millennium) with
specific star formation rates $\log_{10}(SFR/M_\ast)>-11$. Note that in {\tt
Sample IV} 99\% galaxies have specific star fromation rate above this value.
As in the previous figure, the galaxies are divided into two types,
``metal-poor'' and ``metal-rich'' by defining a stellar mass-dependent cut in
exactly the same way as was done for the sample of observed galaxies. As can be
seen, the models do predict a displacement of the metal-poor with respect to
the metal-rich subsamples that is qualitatively reminiscent of the trend seen
in the real data \citep[see also][for a similar finding for 
Milky Way type systems in the Millennium Simulation]{deRossi-2009}. However, the quantitative 
agreement is not so good. In
particular, in the simulations, the two populations are displaced more along
the x-axis (i.e. in stellar mass) than along the y-axis (i.e. in gas fraction),
whereas the opposite is seen in the real data. 

This may indicate the efficiency with which supernovae are assumed to expel gas
from galaxies is too high and that the mass-metallicity trends in the real data
are more driven by variations in cold gas content than in the
\cite{Lucia-Blaizot-07} models. There is a free parameter in the models that
controls the efficiency with which energy from supernova couples to the gas,
and as this increases, the mass-metallicity relation steepens. Variations in
gas fraction are determined by cooling and infall, which should be modeled 
more accurately.

One issue that we have swept under the carpet until now, is that we are
assuming that the total gas content of the galaxy simply scales in proportion
to the H {\sc i} mass, thereby neglecting any intrinsic variations in the ratio
of atomic-to-molecular gas in galaxies. If this ratio were to depend
systematically on metallicity (see for example
\citealt{Krumholz-McKee-Tumlinson-08}), then part of the vertical displacement in
H {\sc i} fraction between metal-poor and metal-rich galaxies may simply be due
to this effect.

\section{Summary}

We have examined the correlation of atomic hydrogen-to-stellar mass ratio,
$G_{HI}/S$, for a sample of 800 galaxies that have H {\sc i} flux measurement
from the HyperLeda catalogues and optical photometry from the SDSS DR4. We use
this sample to derive a new estimator of H {\sc i} mass fraction for
low-redshift galaxies ($z<0.1$). This is $\log_{10}(G_{HI}/S) = -1.73238(g-r) +
0.215182\mu_i - 4.08451$, where $\mu_i$ is the $i$-band surface brightness and ($g-r$)
is the optical colour given by the $g$- and $r$-band Petrosian magnitudes. The
typical scatter in the relation is 0.31 dex. We have tested whether the
residuals in our $G_{HI}/S$ estimator correlate with galaxy properties.  We
find no effect as a function of stellar mass or as a function of mean stellar
age as measured by the 4000 \AA\ break, but there is a small effect as a
function of galaxy concentration. 

We then apply this new estimator to a large sample of star-forming galaxies
from the SDSS DR4 to estimate the H {\sc i} mass function, and we find good
agreement with determinations from recent H {\sc i} surveys. This demonstrates
that our estimator does, at least in a statistical sense, properly reproduce
the distribution of H {\sc i} mass in the local Universe.  We have also used
the data to examine whether the stellar mass-metallicity relation of galaxies
depends on gas content, and we found a systematic change in gas fraction
along this relation. In addition, at fixed stellar mass, galaxies with higher
metallicities tend to contain less gas. 

Finally, we would like to discuss how this work could be improved in future.
First, the calibration sample ({\tt Sample II}) is taken from HyperLeda, which
is an inhomogenous collection of data from different observational programs,
each with different selection effects. It is not clear to what extent our
calibrating sample is fully representative of the local galaxy population. The
fact that we do see some systematic bias as a function of concentration index,
suggests that as the bulge component becomes more prominent, our estimator,
which is motivated by the Kennicutt-Schmidt law for galactic disks, may become
increasingly inaccurate. We intend to test this using the larger and deeper
samples that will soon become available from surveys such as ALFALFA and The GALEX Arecibo SDSS Survey (GASS)
\citep{Catinella-08}.  Another problem is that our
calibrating sample lies at very low redshifts, so that any quantity that is
measured from the SDSS spectra, such as metallicity, is heavily weighted to the
central regions of the galaxies.  We have skirted around this problem by
dividing the galaxy population into two metallicity bins containing equal
number of galaxies and carrying out a {\em relative comparison} between the gas
fractions of the high and low-metallicity sub-samples at fixed stellar mass.
This means that aperture bias are at least roughly the same for our two
sub-samples. Nevertheless, it should be borne in mind that when we refer to
metallicity in this paper, we are not talking about a global-average
quantity.

We believe that it will be most interesting to apply our new indicators to
galaxy samples where there is little prospect getting real H {\sc i} mass
measurements in the foreseeable future, for example at high redshift. The
indicator may also be useful for applications where one would like to have some
rough estimate of the cold gas available to fuel star formation, for example in
satellite galaxies that are destined to merge with their parent object on some
short timescale. We will be looking into such applications in our future work.

\section*{Acknowledgments}

WZ thanks MPA for invitation  and hospitality. We thank the anonymous
referee for helpful comments. This work has been
supported by the Chinese  National Natural Science Foundation  grants 10573020,
10603006, 10633020, 10803007, 10873016 and by National Basic Research Program
of China (973 Program) 2007CB815403. This work has also been supported by the
Young Researcher Grant of National Astronomical Observatories, Chinese Academy
of Sciences, and the Max-Planck Society.

Funding for the SDSS and SDSS-II has been provided by the Alfred P.  Sloan
Foundation, the Participating Institutions, the National Science Foundation,
the U.S. Department of Energy, the National Aeronautics and Space
Administration, the Japanese Monbukagakusho, the Max Planck Society, and the
Higher Education Funding Council for England. The SDSS Web Site is
http://www.sdss.org/.

The SDSS is managed by the Astrophysical Research Consortium for the
Participating Institutions. The Participating Institutions are the American
Museum of Natural History, Astrophysical Institute Potsdam, University of
Basel, University of Cambridge, Case Western Reserve University, University of
Chicago, Drexel University, Fermilab, the Institute for Advanced Study, the
Japan Participation Group, Johns Hopkins University, the Joint Institute for
Nuclear Astrophysics, the Kavli Institute for Particle Astrophysics and
Cosmology, the Korean Scientist Group, the Chinese Academy of Sciences
(LAMOST), Los Alamos National Laboratory, the Max-Planck-Institute for
Astronomy (MPIA), the Max-Planck-Institute for Astrophysics (MPA), New Mexico
State University, Ohio State University, University of Pittsburgh, University
of Portsmouth, Princeton University, the United States Naval Observatory, and
the University of Washington.

\bibliography{sdss,kannappan,kauffmann,himf,mzr,others}

\bsp
\label{lastpage}

\end{document}